\documentstyle[12pt]{article}
\newcommand{\beq}{\begin{equation}}
\newcommand{\eeq}{\end{equation}}
\newcommand{\beqa}{\begin{eqnarray}}
\newcommand{\eeqa}{\end{eqnarray}}
\newcommand{\beqar}{\begin{eqnarray*}}
\newcommand{\eeqar}{\end{eqnarray*}}

\newcommand{\g}{\gamma}
\newcommand{\G}{\Gamma}

\newcommand{\k}{\kappa}

\newcommand{\z}{\zeta}

\newcommand{\eg}{{\it e.g.,}\ }
\newcommand{\ie}{{\it i.e.,}\ }

\newcommand{\pol}{\varepsilon}
\newcommand{\norm}[1]{\raise.3ex\hbox{:}#1\raise.3ex\hbox{:}}
\newcommand{\inn}{\!\cdot\!}
\newcommand{\bz}{\bar{z}}
\newcommand{\bw}{\bar{w}}
\newcommand{\bu}{\bar{u}}
\newcommand{\tV}{\widetilde{V}}

\newcommand{\tS}{\widetilde{S}}
\newcommand{\tX}{\tilde{X}}
\newcommand{\tphi}{\tilde{\phi}}
\newcommand{\tpsi}{\tilde{\psi}}

\newcommand{\labell}[1]{\label{#1}} 
\newcommand{\labels}[1]{\label{#1}}
\newcommand{\reef}[1]{(\ref{#1})}
\newcommand{\slf}{\G} 
\newcommand{\Tr}{{\rm Tr}}

\begin{document}
\begin{titlepage}
\rightline{ \hfill } \vskip 5em

\begin{center}
{\bf \huge Superstring Scattering\\[.25em]
             from O-planes}
\vskip 3em

{\large Mohammad R. Garousi} \vskip 1em

{Department of Physics, Ferdowsi university, P.O. Box 1436, Mashhad, Iran}\\
\vspace*{0.1cm}
and\\
{ Institute for Studies in Theoretical Physics and Mathematics
IPM} \\
{P.O. Box 19395-5531, Tehran, Iran}\\
\vskip 4em

\begin{abstract}
We write the vertex operators of massless  NS-NS and RR states  of
Type II superstring theory in the presence of Orientifold
$p$-planes. They include the usual vertex operators of Type II
theory and their images. We then calculate the two-point functions
of these vertex operators at the projective plane PR$_2$ level. We
show that the result can be written in the Veneziano-type
formulae, with the same kinematic factor that appears in the
D$_p$-branes amplitudes. While the scattering amplitudes with the
usual vertex operators are not gauge invariant, the above
amplitudes are invariant. From the amplitude describing scattering
of two NS-NS states off the O-plane, we find the low energy
effective action of O-planes. The result shows a relative factor
$2^{p-6}$ between couplings to O-planes and to D-branes at
$(\alpha')^2$ order.
\end{abstract}
\end{center}

\end{titlepage}

\setcounter{footnote}{0}
\section{Introduction}
String theory appears to contain various extended objects, other
than just strings. D$_p$-branes, for instance,  are dynamical
extended objects that have remarkably simple word sheet
description. World-sheet with boundary on which spacetime
coordinate fields satisfy Dirichlet and Neumann boundary
conditions. Using this simple description, it was shown in
\cite{joe} that D-branes carry Ramond-Ramond (RR) charges in type
II superstring theory. Another set of extended objects which have
simple world sheet description are Orientifold $p$-planes. These
are non-dynamical planes at the fixed points of spacetime. The
world-sheet in this case has crosscap instead of boundary. Having
world sheet description, one can study these O-planes using well
understood tools of two dimensional quantum field theory.
Interaction between O-plane and D-brane, for instance, has been
found. From this amplitude one finds that O-planes have negative
tension, and carry RR charge in the case of Type II superstring
theory (see \eg \cite{Dbreview}). We are interested, in this
paper, in studying closed string scattering with O-planes. Closed
string scattering off O-planes have been studied in \cite{KD,HJS}
using boundary state formalism. However, we would like to study
them here using projective plane calculation.



In string theory, the description for scattering of closed string
off D$p$-brane is as follows: The string background is taken to be
simply flat empty space, however, interactions of closed strings
with a D$p$-brane are described by world-sheets with boundary.
The boundary of world sheet must be fixed to the  surface at the
position of the D$p$-brane. The latter is accomplished by imposing
Neumann boundary conditions on the fields along D$_p$-brane world
volume, and Dirichlet boundary conditions on the  fields
transverse to the D$_p$-brane \cite{Dbrane,Dbreview}.
The scattering amplitudes describing the
interactions of closed strings with D-branes has been found in
\cite{igora,igorb,form,MG}, \ie \beqa
A^{D_2}&\simeq&K(1,2)\frac{\G(-t/2)\G(-2s)}{\G(1-t/2-2s)}\nonumber\eeqa
where $K(1,2)$ is kinematic factor that depends on momentum and
polarization of external states. It satisfies various Ward
identities. In this equation, $t$ is the closed string channel and
$s$ is open string channel.

The description for scattering of closed string off O$p$-plane is
as follows: The string background is again taken to be  flat empty
space,  however, interactions of closed strings with these
O$p$-planes are described by world-sheets with a crosscap instead
of boundary.
The crosscap must be fixed to the surface at the position of the
O$p$-plane. The latter is accomplished by choosing appropriate
boundary conditions \cite{CC}.
The scattering amplitude of closed string off O-plane has been
studied in \cite{KD,HJS} using the boundary states formalism. In
these studies, the standard vertex operator of closed string
states, \ie the same vertex operator that one uses in scattering
with D-brane, have been used. In \cite{HJS}, it has been argued
that the scattering of NS-NS closed string off O-plane  is not
gauge invariant because the standard vertex operators have been
used in calculating the amplitude.  We note here that in the
presence of O-plane, the vertex operators must satisfy the
constraint that when they are on the world volume of O-plane they
should describe unoriented states. Using this constraint, we will
find the vertex operators in the presence of O-plane and will find
that the scattering amplitudes describing the interactions of
closed strings with O$_p$-planes are given by \beqa
A^{PR_2}&\simeq&K(1,2)\frac{\G(-t/2)\G(-u/2)}{\G(1-t/2-u/2)}\labell{zero}\eeqa
where $K(1,2)$ is ${\it exactly}$ the same kinematic factor that
appears in the D-brane case. Hence, the amplitude is gauge
invariant. In this amplitude, both channels are closed string
channel. Here $t=-(p_1+p_2)^2$ and $u=-(p_1+p_2\inn D)^2$ where
$p_i$ is momentum of external state and $p_i\inn D$ is momentum of
its image.

 The paper is organized as
follows: In the following section, after writing the NS-NS vertex
operator in the presence of O-plane,  we describe the calculation
of the scattering of two such  states  from an O-plane using
conformal field theory techniques. The result is  fully covariant,
satisfy the Ward identity and can be written as \reef{zero}. In
section 3, we repeat  the same calculation for scattering
amplitude of two RR states from O-plane. In this case also we show
that the result can be written as \reef{zero}.
We then examine the NS-NS  amplitude at low energy in section 4.
We show in this section that the massless poles of the amplitude
are in exact agreement with    analogous field theory
calculations, the contact terms at $(\alpha')^0$ order are
consistent with negative cosmological constant of O-plane, and the
effective action at $(\alpha')^2$ order is the same as the
effective action of D-brane with the relative factor $2^{p-6}$.
We conclude with a discussion of our results in section 5.

\section{NS-NS scattering amplitudes} \labels{nsns}

T-duality of unoriented Type I theory gives Type II theory with
O-planes at the fixed points of the T-dual space. In the original
Type I theory, states are invariant under  world-sheet parity
$\Omega$. In the T-dual theory, this operator has two effects,
world sheet parity and spacetime reflection on the T-dualized
coordinates $x^i$. If one considers only the center of mass
dependence of the string wave function, then the projection onto
$\Omega=+1$ determines the wave function at $-x^i$ to be the same
as that at $x^i$, up to a sign\footnote{Our index conventions are
that lowercase Greek indices take values in the entire
ten-dimentional spacetime, \eg $\mu\nu=0,1,...,9$; early Latin
indices take  values in the world volume, \eg $a,b,c=0,1,...,p$;
and middle Latin indices take values in the transverse space, \eg
$i,j=p+1,...,9$. Thus, for example, $G_{\mu\nu}$ denotes the
entire spacetime metric, while $G_{ab}$ and $G_{ij}$ denote metric
components for directions parallel and orthogonal to the
O-planes/D-branes, respectively.}. Various components of massless
states in NS-NS sector satisfy \cite{polch}\beqa
G_{ab}(x^a,-x^i)&=&G_{ab}(x^a,x^i)\qquad ; \qquad
B_{ab}(x^a,-x^i)=-B_{ab}(x^a,x^i)\nonumber\\
 G_{ai}(x^a,-x^i)&=&-G_{ai}(x^a,x^i)\qquad ; \qquad
B_{ai}(x^a,-x^i)=B_{ai}(x^a,x^i)\nonumber\\
G_{ij}(x^a,-x^i)&=&G_{ij}(x^a,x^i)\qquad ; \qquad
B_{ij}(x^a,-x^i)=-B_{ij}(x^a,x^i)\eeqa The orientifold fixed plane
is at $x^i=0$. The wave functions that satisfy the above
conditions are \beqa
G_{ab}&=&\pol_{ab}e^{ip_ax^a}\cos(p_ix^i)\qquad;\qquad
B_{ab}=i\pol_{ab}e^{ip_ax^a}\sin(p_ix^i)\nonumber\\
G_{ai}&=&i\pol_{ai}e^{ip_ax^a}\sin(p_ix^i)\qquad;\qquad
B_{ai}=\pol_{ai}e^{ip_ax^a}\cos(p_ix^i)\nonumber\\
G_{ij}&=&\pol_{ij}e^{ip_ax^a}\cos(p_ix^i)\qquad;\qquad
B_{ij}=i\pol_{ij}e^{ip_ax^a}\sin(p_ix^i) \eeqa where
$\pol_{\mu\nu}$ is the polarization tensor. The momentum and
polarization tensor satisfy $p^2=0$, and
 $p^\mu\,\pol_{\mu\nu}=0=\pol_{\mu\nu}\,p^\nu $. One can rewrite the
above wave functions as \beqa
G_{\mu\nu}&=&\frac{1}{2}\left(\pol_{\mu\nu}e^{ip\cdot
x}+(D\inn\pol\inn D)_{\mu\nu}e^{ip\cdot D\cdot
x}\right)\nonumber\\
B_{\mu\nu}&=&\frac{1}{2}\left(\pol_{\mu\nu}e^{ip\cdot
x}-(D\inn\pol\inn D)_{\mu\nu}e^{ip\cdot D\cdot
x}\right)\labell{polg}\eeqa where matrix $D$ is $\eta_{ab}$ for
world volume directions and is $-\eta_{ij}$  for transverse
directions. The vertex operator corresponding to the above wave
functions is \beqa V^{PR_2}(\pol,p,z,\bz)&\!\!\!=\!\!\!&
\frac{1}{2}\left(\pol_{\mu\nu}\norm{
V^{\mu}_{\alpha}(p,z)}\norm{\tV^{\nu}_{\beta}(p,\bz)}+(D\inn\pol^T\inn
D)_{\mu\nu}\norm{V^{\mu}_{\alpha}(p\inn
D,z)}\norm{\tV^{\nu}_{\beta}(p\inn
D,\bz)}\right)\labell{ver1}\eeqa The holomorphic components
$V^{\mu}_{\alpha}(p,\pol)$ in 0 and -1 picture are given by \beqa
V_{-1}^\mu(p,z)&=&e^{-\phi(z)}\,\psi^{\mu}(z)\,e^{ip\cdot X(z)}
\nonumber\\
V_0^\mu(p,z)&=&\left(\partial X^\mu(z)+ip\inn \psi(z)\psi^{\mu}(z)
\right)\,e^{ip\cdot X(z)}\ \ . \labell{vertright} \eeqa The
antiholomorphic components take the same form as in
eq.~(\ref{vertright}) but with the left-moving fields replaced by
their right-moving counterparts -- \ie $X(z)\rightarrow\tX(\bz)$,
$\psi(z)\rightarrow\tpsi(\bz)$, and
$\phi(z)\rightarrow\tphi(\bz)$.
The vertex operator \reef{ver1} can also be
written as \beqa V^{PR_2}(\pol,p,z,\bz)&\!\!\!=\!\!\!&
\frac{1}{2}\left(\pol_{\mu\nu}\norm{
V^{\mu}_{\alpha}(p,z)}\norm{\tV^{\nu}_{\beta}(p,\bz)}\right.\labell{ver2}\\
&&\left.+\pol_{\mu\nu}\left(\norm{V^{\mu}_{\alpha}(p,z)}\norm{\tV^{\nu}_{\beta}(p,\bz)}\right)
_{\psi(z)\leftrightarrow D\cdot\tpsi(\bz),\,X(z)\leftrightarrow
D\cdot\tX(\bz),\,\phi(z)\leftrightarrow
\tphi(\bz)}\right)\nonumber\eeqa For the special case that
$D^{\mu\nu}=\eta^{\mu\nu}$ \ie no T-duality on Type I theory,  one
should recover the NS-NS vertex operator of unoriented Type I
theory. Interchanging the holomorphic part with the
antiholomorphic part in the second line above, one finds that the
two lines are the same except the fact that the transpose of
polarization appears in the second line. Hence, the Kalb-Ramond
antisymmetric tensor is projected out. The surviving states are
graviton and dilaton which are the only NS-NS states in unoriented
Type I theory. Note also that on the O$_p$-plane world volume
$X^i=0=\psi^i$. Hence,  only matrix $D^{ab}=\eta^{ab}$ appears in
the vertex operator. As a result, as expected, the vertex operator
represents  unoriented states on the world volume of O-plane.

For D-brane and in the absence of O-planes, there is no constraint
on the closed string states, hence, the  wave function for
graviton/antisymmetric tensor is\beqa \pol_{\mu\nu}e^{ip\cdot
x}\labell{waved}\eeqa and the corresponding vertex operator is
twice the first term in \reef{ver1}.

The  amplitudes describing the scattering of two massless
\hbox{NS-NS} states from an O-plane/D-brane  are calculated as two
closed string vertex operator insertions on a PR$_2$/D$_2$ with
appropriate crosscap/boundary conditions \cite{Dbrane,Dbreview}.
The amplitude is given  as \beq A\simeq\int d^2\!z_1\, d^2\!z_2\
\langle\, V_1(\pol_1,p_1,z_1,\bz_1) \
V_2(\pol_2,p_2,z_2,\bz_2)\,\rangle \labell{ampone} \eeq
both integrals run over the unit disk. The choices for the picture
of vertex operators are arbitrary, however, the sum of superghost
charge must be -2. The amplitude can be separated as \beqa
A^{PR_2}&=&A_1+A_2+A_3+A_4\nonumber\\
A^{D_2}&=&A_1\eeqa where \beqa A_1&\!\!\!\simeq\!\!\!&
\frac{1}{4}\pol_{1\mu\nu}\pol_{2\alpha\beta}\int
d^2\!z_1\,d^2\!z_2\langle\,\norm{V^{\mu}_{-1}(p_1,z_1)}\,\norm{\tV^{\nu}_0(p_1,\bz_1)}\,
\norm{V^{\alpha}_{-1}(p_2,z_2)}\,\norm{\tV^{\beta}_0(p_2,\bz_2)}\rangle\labell{amp1}\\
A_2&\!\!\!\simeq\!\!\!&
\frac{1}{4}\pol_{1\mu\nu}(D\inn\pol_2^T\inn D)_{\alpha\beta}\int
d^2\!z_1\,d^2\!z_2\langle\,\norm{V^{\mu}_{-1}(p_1,z_1)}\,\norm{\tV^{\nu}_0(p_1,\bz_1)}\,
\norm{V^{\alpha}_{-1}(p_2\inn D,z_2)}\,\norm{\tV^{\beta}_0(p_2\inn D,\bz_2)}\rangle\nonumber\\
A_3&\!\!\!\simeq\!\!\!& \frac{1}{4}(D\inn\pol_1^T\inn
D)_{\mu\nu}\pol_{2\alpha\beta}\int
d^2\!z_1\,d^2\!z_2\langle\,\norm{V^{\mu}_{-1}(p_1\inn D
,z_1)}\,\norm{\tV^{\nu}_0(p_1\inn D,\bz_1)}\,
\norm{V^{\alpha}_{-1}(p_2,z_2)}\,\norm{\tV^{\beta}_0(p_2,\bz_2)}\rangle\nonumber\\
A_4&\!\!\!\simeq\!\!\!&\frac{1}{4} (D\inn\pol_1^T\inn
D)_{\mu\nu}(D\inn\pol_2^T\inn D)
_{\alpha\beta}\nonumber\\
&&\qquad\qquad\times\int
d^2\!z_1\,d^2\!z_2\langle\,\norm{V^{\mu}_{-1}(p_1\inn D
,z_1)}\,\norm{\tV^{\nu}_0(p_1\inn D,\bz_1)}\,
\norm{V^{\alpha}_{-1}(p_2\inn D
,z_2)}\,\norm{\tV^{\beta}_0(p_2\inn D,\bz_2)}\rangle\nonumber
\eeqa To calculate the above correlators, one needs propagator for
various world-sheet fields.
Separately,  the left- and right-moving fields have  standard
propagators on sphere world sheet, \eg \beqa \langle
X^{\mu}(z)\,X^{\nu}(w)\rangle&=&-\eta^{\mu \nu}\,\log(z-w)
\nonumber\\
\langle\psi^{\mu}(z)\,\psi^{\nu}(w)\rangle&=&-\frac{\eta^{\mu
\nu}}{z-w}
\nonumber\\
\langle\phi(z)\,\phi(w)\rangle&=&-\log(z-w) \labell{standard}
\eeqa with analogous expressions for the right-movers (see
\cite{danf}). However, the projective plane/disk can be obtained
as a $Z_2$ identification of the sphere. This identifies points
$z$ and $z'$ with $z'=-n/\bz$ where $n=1$ for projective plane and
$n=-1$ for disk. The method of images can then be used to obtain
the propagators between left- and right-moving fields
\cite{polch1}. Alternatively, the boundary state of O-plane is
obtained by inserting $(-1)^m$ in every terms of the exponent of
the D-brane boundary state \cite{CC}. This means every $z_i\bz_j$
combination of D-brane gets replaced by $-z_i\bz_j$ \cite{KD}. In
the propagator between left- and right-moving fields each $z\bw$
should be replaced by $-z\bw$. Hence,  \beqa \langle
X^{\mu}(z)\,\tX^{\nu}(\bw)\rangle&=&-D^{\mu \nu}\,\log(1+nz\bw)
\labell{xcor}\\
\langle\psi^{\mu}(z)\,\tpsi^{\nu}(\bw)\rangle&=&-\frac{\sqrt{n}D^{\mu\nu}}{1+nz\bw}\,;\,\,\,\,\,\,
\langle\tpsi^{\mu}(\bw)\,\psi^{\nu}(z)\rangle\,=\,\frac{\sqrt{n}D^{\mu\nu}}{1+nz\bw}
\nonumber\nonumber\\
\langle\phi(z)\,\tphi(\bw)\rangle&=&-\log(1+nz\bw)\ \ . \nonumber
\eeqa For D$_p$-brane case, the fields $X^a$  for $a=0,1,\cdots ,
p$ should satisfy  Neumann boundary conditions, and $X^i$ for
$i=p+1,\ldots,9$ should satisfy the Dirichlet boundary conditions
\cite{Dbrane,Dbreview}.
These facts are incorporated by the matrix $D^{\mu\nu}$. In the
scattering amplitude, this matrix has  also the effect of having
conservation of momentum along the world volume of D$_p$-brane. To
obtain $PR_2$ propagators, we  have replaced $z\bw$ by $-z\bw$.
In the above propagators, we have fixed the coefficient of the
propagator of world sheet fermion  by requiring  that the
coefficient of unphysical vertex operator in the OPE of
$V^{\mu}_0(p_1,z_1)$ and $\tV^{\nu}_0(p_2,\bz_2)$ must be a total
derivative \cite{MBG}, \ie \beqa
:V^{\mu}_0(p_1,z_1)::\tV^{\nu}_0(p_2,\bz_2):&\simeq&n\eta^{\mu\nu}\frac{p_1\inn
D\inn p_2-1}{(1+nz_1\bz_2)^{p_1\cdot D\cdot p_2-2}}e^{ip_1\cdot
X+ip_2\cdot\tX}+\cdots\nonumber\eeqa

Given  the above propagators, it is now straightforward to
evaluate the correlation functions appearing in the amplitude
(\ref{amp1}). After evaluating these correlators, one finds \beqa
A_1&\simeq&\frac{1}{4}\pol_1^{\mu\nu}\pol_2^{\alpha\beta}\int
d^2z_1d^2z_2|z_1-z_2|^{2p_1\cdot p_2}|1+nz_1\bz_2|^{2p_1\cdot
D\cdot p_2}(1+nz_1\bz_1)^{p_1\cdot D\cdot
p_1}(1+nz_2\bz_2)^{p_2\cdot D\cdot p_2}\nonumber\\
&&\qquad\qquad\times\left(I^1_{\mu\nu\alpha\beta}+I^2_{\mu\nu\alpha\beta}
+I^3_{\mu\nu\alpha\beta}+I^4_{\mu\nu\alpha\beta}\right)\delta^{p+1}(p_1+p_1\inn
D+p_2+p_2\inn D)\labell{amp01}\eeqa where delta function gives
conservation of momentum along world volume of O-plane/D-brane.
There is a factor of
$i(2\pi)^{p+1}C^X_{PR_2}C^{\psi}_{PR_2}C^{\phi}_{PR_2}$ for $PR_2$
case and $i(2\pi)^{p+1}C^X_{D_2}C^{\psi}_{D_2}C^{\phi}_{D_2}$ for
disk where $C^X_{PR_2}$, for example,  is the functional
determinant for the string field $X^a$  \cite{polch1}. We are not
trying to calculate these constants here. Hence, we omit them here
and at the end normalize the amplitude by comparing with low
energy effective action. In the above equation,\beqa
I^1_{\mu\nu\alpha\beta}&=&\frac{\eta_{\mu\alpha}}{(z_1-z_2)^2}\left(\frac{\eta_{\nu\beta}}{(\bz_1-\bz_2)^2}
+\left[-\frac{p_1\inn D_{\nu}nz_1}{(1+nz_1\bz_1)}-\frac{p_2\inn
D_{\nu}nz_2}{(1+nz_2\bz_1)}+\frac{p_{2\nu}}{(\bz_2-\bz_1)}\right]\right.\nonumber\\
&&\qquad\qquad\qquad\left.\times \left[-\frac{p_1\inn
D_{\beta}nz_1}{(1+nz_1\bz_2)}-\frac{p_2\inn
D_{\beta}nz_2}{(1+nz_2\bz_2)}+\frac{p_{1\beta}}{(\bz_1-\bz_2)}\right]\right)\nonumber\\
I^2_{\mu\nu\alpha\beta}&=&\frac{n}{(z_1-z_2)} \left[-\frac{p_1\inn
D_{\beta}nz_1}{(1+nz_1\bz_2)}-\frac{p_2\inn
D_{\beta}nz_2}{(1+nz_2\bz_2)}+\frac{p_{1\beta}}{(\bz_1-\bz_2)}\right]\nonumber\\
&&\qquad\qquad\qquad\times \left(\frac{p_1\inn
D_{\mu}D_{\nu\alpha}-p_1\inn
D_{\alpha}D_{\mu\nu}}{(1+nz_1\bz_1)(1+nz_2\bz_1)}\right)\nonumber\\
I^3_{\mu\nu\alpha\beta}&=&\frac{n}{(z_1-z_2)} \left[-\frac{p_1\inn
D_{\nu}nz_1}{(1+nz_1\bz_1)}-\frac{p_2\inn
D_{\nu}nz_2}{(1+nz_2\bz_1)}+\frac{p_{2\nu}}{(\bz_2-\bz_1)}\right]\nonumber\\
&&\qquad\qquad\qquad\times \left(\frac{p_2\inn
D_{\mu}D_{\alpha\beta}-p_2\inn
D_{\alpha}D_{\mu\beta}}{(1+nz_1\bz_2)(1+nz_2\bz_2)}\right)\nonumber\\
I^4_{\mu\nu\alpha\beta}&=&\frac{n}{(z_1-z_2)} \left(\frac{p_1\inn
D_{\mu}(\eta_{\nu\beta}p_2\cdot
D_{\alpha}-p_{2\nu}D_{\alpha\beta})-D_{\mu\nu}(p_2\cdot
D_{\alpha}p_{1\beta}-p_1\cdot p_2 D_{\alpha\beta})
}{(1+nz_1\bz_1)(\bz_1-\bz_2)(1+nz_2\bz_2)}\right.\nonumber\\
&&+\left.\frac{p_2\inn D_{\mu}(-\eta_{\nu\beta}p_1\cdot
D_{\alpha}+p_{1\beta}D_{\alpha\nu})+D_{\mu\beta}(p_1\cdot
D_{\alpha}p_{2\nu}-p_1\cdot p_2 D_{\alpha\nu})
}{(1+nz_1\bz_2)(\bz_1-\bz_2)(1+nz_2\bz_1)}\right.\nonumber\\
&&\qquad\qquad\qquad\left.-n\frac{\eta_{\mu\alpha}(p_1\cdot
p_2\eta_{\nu\beta}-p_{1\beta}p_{2\nu})}{(z_1-z_2)(\bz_1-\bz_2)^2}\right)\labell{I1234}
\eeqa

Now the integrand should be invariant under the CKG of the
projective plane/disk which is $SL(2,C)$ group of sphere that
respects the identification $z=-n/\bz$.
Transformation of $z$ and $\bz$ under this group are \beqa
z\rightarrow \frac{az+b}{cz+d}\,\,&;&\,\,\bz\rightarrow
-n\frac{d\bz-nc}{b\bz-na}\eeqa where $a,b,c,d$ are four complex
parameters that satisfy $ad-cb=1$. Moreover, comparing the second
transformation above with the complex conjugate of first
transformation gives $\bar{b}=-nc$ and $\bar{d}=a$. Under above
transformation, one finds\beqa
z_1-z_2&\rightarrow&\frac{z_1-z_2}{(cz_1+d)(cz_2+d)}\nonumber\\
\bz_1-\bz_2&\rightarrow&\frac{\bz_1-\bz_2}{(b\bz_1-na)(b\bz_2-na)}\nonumber\\
1+nz_1\bz_2&\rightarrow&\frac{1+nz_1\bz_2}{(cz_1+d)(b\bz_2-na)}\nonumber\\
dzd\bz&\rightarrow&\frac{dzd\bz}{(cz+d)^2(b\bz-na)^2} \eeqa One
can  check that the integrand in the amplitude \reef{amp01} is
invariant under the above transformations. In checking this, one
should replace one of the momentum in each bracket $[...]$ in
\reef{I1234} in terms of other two momentum using conservation of
momentum and the physical  condition on the polarization of
external states, \ie $p_i\inn\pol_i=0$. To gauge fix this
symmetry, we first fix one vertex operator at $z_1=\bz_1=0$. The
residual symmetry is then $z\rightarrow \frac{a}{\bar{a}} z$ which
is a rotation in the $z$-plane. To fix the residual symmetry, we
fix the polar coordinate of the second vertex operator, \ie
$|z_2|=|\bz_2|=r$. Ignoring the volume of the symmetry group, \ie
\beqa d^2\!z_1\ d^2\!z_2&\rightarrow &dr^2\labell{mesure} \eeqa
one is  left with a single real integral of the form \beqa
A^{D_2}&=&\frac{i\kappa T_p}{2}\,  \int_0^1dr^2\ (r^2)^{p_1\cdot
p_2} (1-r^2)^{p_2\cdot D\cdot p_2}
\left[\frac{a_1}{r^2}+\frac{a_2}{1-r^2}+\frac{a_3}{r^4}\right]
\labell{norm}\nonumber\\
A^{PR_2}_1&=&\frac{i\kappa T'_p}{8}\,  \int_0^1dr^2\
(r^2)^{p_1\cdot p_2} (1+r^2)^{p_2\cdot D\cdot p_2}
\left[\frac{a_1}{r^2}+\frac{a_2}{1+r^2}+\frac{a_3}{r^4}\right]
\eeqa where $a_1,\, a_2$ and $a_3$ are three kinematic factors
depending only on the spacetime momenta and polarization tensors.
We have also normalized the amplitudes at this point by the
introduction of factors of $\kappa$, the closed string coupling,
and $T'_p(T_p)$, the O-plane (D-brane) tension,
respectively.\footnote{Here and in the subsequent amplitudes, we
omit the Dirac delta-function which imposes momentum conservation
in the  world-volume directions. We have introduced a phase $-i$
though which corresponds to that of the analogous field theory
amplitudes calculated in Minkowski space
--- see sect.4.} Note that the constants that we have omitted in
\reef{amp01} and in \reef{mesure} have the same sign for both
$PR_2$ and $D_2$. Hence, the constants $T_p'$ and $T_p$ must have
the same sign. The kinematic factors are\beqa
a_3&=&\Tr(\pol_1\inn \pol_2^T)(1+t/2)\nonumber\\
 a_1&=&n\left({\rm
Tr}(\pol_1\inn D)\,p_1\inn \pol_2 \inn p_1 -p_1\inn\pol_2\inn
D\inn\pol_1\inn p_2 - p_1\inn\pol_2\inn\pol_1^T \inn D\inn
p_1\right.
\nonumber\\
&&\left.\ -p_1\inn\pol_2^T \inn \pol_1 \inn D \inn p_1 -
\frac{1}{2}(p_2\inn\pol_1^T\inn\pol_2 \inn p_1+
p_1\inn\pol_2\inn\pol_1^T \inn p_2)\right.\nonumber\\
&&\left.- \frac{t}{4}\left(\Tr(\pol_1\inn D)\Tr(\pol_2\inn
D)-\Tr(\pol_1\inn D\inn\pol_2\inn D)\right)
+\Big\{1\longleftrightarrow
2\Big\}\right)\nonumber\\
a_2&=&{\rm Tr}(\pol_1\inn D)\,(p_1\inn\pol_2\inn D\inn p_2 -
p_2\inn D\inn\pol_2\inn D\inn p_1 )
\nonumber\\
&&+p_1\inn D\inn\pol_1\inn D\inn\pol_2\inn D\inn p_2 -p_2\inn
D\inn\pol_2\inn\pol_1^T\inn D\inn p_1
\nonumber\\
&&+{\rm Tr}(\pol_1\inn D) {\rm Tr}(\pol_2\inn D)\,(t/4)
+\Big\{1\longleftrightarrow 2 \Big\} \labell{fintwo} \eeqa where
$t=-(p_1+p_2)^2=-2p_1\inn p_2$ is the momentum transfer to the
O-plane/D-brane, and $s=-p_1^a p_1^b\eta_{ab}=-{1\over2}p_1\inn
D\inn p_1$ is the momentum flowing parallel to the world-volume of
the O-plane/D-brane. Our notation is such that \eg
$p_1\inn\pol_2\inn\pol_1^T \inn D\inn p_1
=p_1^\mu\,\pol_{2\mu\nu}\,\pol_1{}^{\lambda\nu}\,D_{\lambda\rho}\,p_1^\rho$.

Using the following integrals:\beqa
\int_0^1dx\,x^{-\alpha}(1+x)^{-\beta}&=&\frac{F(\beta,1-\alpha;2-\alpha;-1)}{1-\alpha}\nonumber\\
\int_0^1dx\,x^{-\alpha}(1-x)^{-\beta}&=&B(1-\alpha,1-\beta) \eeqa
One finds \beqa A^{D_2}&=&\frac{i\kappa\,
T_p}{2}\left(a_3B(-1-t/2,1-2s)\right.\labell{finone}\\
&&\qquad\qquad\qquad \left.+ a_1B(-t/2,1-2s)+
a_2B(1-t/2,-2s)\frac{}{}\right)\nonumber\\
A^{PR_2}_1&=&\frac{i\kappa\,T'_p}{8}\left(a_3\frac{F(2s,-1-t/2;-t/2;-1)}{-1-t/2}\right.\labell{finone2}\\
&&\qquad\qquad\qquad \left.+a_1\frac{F(2s,-t/2;1-t/2;-1)}{-t/2}
+a_2\frac{F(2s+1,1-t/2;2-t/2;-1)}{1-t/2} \right) \nonumber\eeqa As
is evident the final amplitude is symmetric under the interchange
of the two string states, \ie $1\longleftrightarrow 2$, despite
the asymmetric appearance of the initial integrand in
eq.~(\ref{ampone}). From the gamma function factors appearing
$D_2$ amplitude, we see that the amplitudes contain two infinite
series poles\footnote{We explicitly restore $\alpha^\prime$ here.
Otherwise our conventions set $\alpha^\prime=2$.} corresponding to
closed string states in the $t$-channel with $\alpha^\prime
m^2=4n$, and to open string states in the $s$-channel with
$\alpha^\prime m^2=n$, with $n=0,1,2,\ldots$. However, the $PR_2$
amplitude has
 poles only in the $t$-channel with $\alpha^\prime m^2=4n$.
 Having no open
string pole is consistent with the fact that the $O_p$-plans,
 unlike the $D_p$-branes,  are not
dynamical objects. The amplitude $A^{D_2}$ is the one that has
been found in \cite{MG} using upper-half z-plane calculation.
While this amplitude satisfies the Ward identities associated with
the gauge invariance of external states, \ie the amplitude
vanishes upon substituting $\pol_{i\mu\nu}\rightarrow p_{i\mu}\,
q_{i\nu}$ or $q_{i\mu}\, p_{i\nu}$, where $q_i\cdot p_i=0$
\cite{MG}, the amplitude $A_1^{PR_2}$ does not satisfy the Ward
identities. Using the package \cite{TH} for expanding the
hypergeometric functions, one finds that $A_1^{PR_2}$ has the
following expansion at low energy\footnote{I would like to thank
H. Ghorbani for finding this expansion.}: \beqa
A_1^{PR_2}&=&\frac{i\k
T'_p}{8}\left(-\frac{1}{t}\left[(t-4s)\Tr(\pol_1\inn\pol_2^T)+2a_1\right]\right.\nonumber\\
&&+\ln(2)\left[4s\Tr(\pol_1\inn\pol_2^T)+a_2\right]\nonumber\\
&&+\frac{\pi^2}{24}\left[2s(t+4s)\Tr(\pol_1\inn\pol_2^T)-4sa_1+ta_2\right]-
\left(\ln(2)\right)^2\left[4s^2\Tr(\pol_1\inn\pol_2^T)+sa_2\right]\nonumber\\
&&\left.+\cdots \frac{}{}\right)\eeqa The massless pole is
reproduced  in field theory assuming linear coupling of gravity to
O-plane. However, the $(\alpha')^0$ contact term is not consistent
with the cosmological constant  of the O-plane effective action.
Moreover, the $\alpha'$ and $(\alpha')^2$ contact terms above are
not consistent with what is expected for the effective action of
O-plane.  This part of amplitude  has been also found in
\cite{HJS} using the boundary state formalism. However, they have
used another picture for the graviton vertex operator, \ie they
have used $V_{(0,0)}$ whereas we use $V_{(-1,0)}$. Their  result
is the same as $A_1^{PR_2}$ above up to some contact terms. It has
been argued in \cite{HJS} that the $\ln(2)$ terms should be absent
in the final gauge invariant amplitude since their presence would
be in conflict with the fact that such terms are not present in
the effective action of type I string theory. Using this, an
expansion for the amplitude $A_1^{PR_2}$ has been found in
\cite{HJS} that is consistent with expected low energy effective
action. We will show shortly, however,  that when we add the
contributions from $A_2$, $A_3$ and $A_4$ the final result
satisfies the Ward identities, it does not contain the $\ln(2)$
terms, it is independent of the choice for the picture of vertex
operators, and it produces expected low energy effective action.

Before calculating $A_2,\, A_3,\, A_4$, we would like to compare
the t-channel poles in  the  amplitudes \reef{finone}  and
\reef{finone2}. The Feynman rule for calculating the t-channel
poles in field theory is as follows: It is a summation of infinite
terms. Each is a product of a vertex of three states where two of
them are the on-shell external states, the propagatore of the
off-shell state, and the coupling of the off-shell state to
O-plane/D-brane. Using the fact that the external states  are the
same in both amplitudes, the t-channel analysis gives information
about the linear coupling of off-shell closed string states to
O-plane and to D-brane. To compare them,
 consider the
following definition of  the beta function (see \eg \cite{gsw}):
\beqa
B(\alpha,\beta)&=&\sum_{n=0}^{\infty}\frac{(-1)^n}{n!}\frac{(\beta-1)(\beta-2)\cdots
(\beta-n)}{\alpha+n}\labell{expand1}\eeqa This expansion can be
rewritten as \beqa
B(\alpha,\beta)&=&\frac{1}{\Gamma(1-\beta)}\sum_{n=0}^
{\infty}\frac{\Gamma(n+1-\beta)\Gamma(n+\alpha)}{\Gamma(n+\alpha+1)}\nonumber\\
&=&\frac{1}{\alpha}F(1-\beta,\alpha;1+\alpha,1)\eeqa From this,
one finds the following expansion for the hypergeometric function
appearing in \reef{finone2}: \beqa
\frac{1}{\alpha}F(1-\beta,\alpha;1+\alpha,-1)&=&
\sum_{n=0}^{\infty}\frac{1}{n!}\frac{(\beta-1)(\beta-2)\cdots
(\beta-n)}{\alpha+n}\labell{expand2}\eeqa Using \reef{expand1} and
\reef{expand2}, one finds the $t$-channel expansion for the
amplitudes in \reef{finone} and \reef{finone2}. Noting  that
$a_1^{D_2}=-a_1^{PR_2}$, one finds that the two amplitudes have
the same pole structure, however, the overall sign of each pole in
two amplitudes is different . The signs are such that the massless
poles have opposite sign, the first massive poles have the same
sign, the second massive poles have again opposite sign, and so
on.
This observation can also be made  from the interaction of D-brane
with O-plane \cite{Dbreview}.

Now to calculate $A_2,\, A_3,\, A_4$, one may try to find them
from $A_1$ by replacing appropriate momenta and polarization
tensors, \eg if $(p_1,\pol_1,p_2,\pol_2)\rightarrow
(p_1,\pol_1,p_2\inn D, D\inn\pol_2^T\inn D)$, then $A_1\rightarrow
A_2$. In this way, using the identity $(D\inn
D)^{\mu\nu}=\eta^{\mu\nu}$, one can easily show that \beqa
A_1&=&A_4\qquad;\qquad A_2=A_3\eeqa Finding $A_2$ in this way, the
final result, $A^{PR_2}$, would be in terms of some hypergeometric
functions which we have found  it difficult to simplify the final
result. Instead, we calculate $A_2$ using another arrangement  for
the picture of the vertex operators in \reef{amp1}. We write it
as\beqa A_2&\!\!\!\simeq\!\!\!& \pol_{1\mu\nu}(D\inn\pol_2\inn
D)_{\alpha\beta}\int
d^2\!z_1\,d^2\!z_2\langle\,\norm{V^{\mu}_{-1}(p_1,z_1)}\,\norm{\tV^{\nu}_0(p_1,\bz_1)}\,
\norm{V^{\beta}_{0}(p_2\inn
D,z_2)}\,\norm{\tV^{\alpha}_{-1}(p_2\inn
D,\bz_2)}\rangle\nonumber\eeqa After evaluating the correlators
and fixing the CKG symmetry as before, one will find the following
result: \beqa A^{PR_2}_2&=&\frac{i\kappa T'_p}{8}\, \int_0^1dr^2\
(r^2)^{p_1\cdot D\cdot p_2} (1+r^2)^{p_2\cdot D\cdot p_2}
\left[\frac{a_1}{r^2}+\frac{a_2}{r^2(1+r^2)}+a_3\right]\eeqa where
$a_1$, $a_2$, and $a_3$ are the kinematic factors \reef{fintwo}.
In terms of the hypergeometric function it is \beqa
A_2&=&\frac{i\kappa\,T'_p}{8}\left(a_3\frac{F(2s,1-u/2;2-u/2;-1)}{1-u/2}\right.\nonumber\\
&&\qquad\qquad\qquad \left.+a_1\frac{F(2s,-u/2;1-u/2;-1)}{-u/2}
+a_2\frac{F(2s+1,-u/2;1-u/2;-1)}{-u/2} \right) \nonumber\eeqa
where $u=-(p_1+p_2\inn D)^2=-2p_1\inn D\inn p_2$. This amplitude
has an infinite tower of closed string poles in $u$-channel, and
has no open string $s$-channel. This amplitude like  $A_1^{PR_2}$
in \reef{finone} does not satisfy the Ward identities. However, we
will see shortly that the combination of the two parts does.
Considering  the  following identity \cite{wolf}: \beqa
F(a,b;b+1;-1)&=&\frac{b}{b-a}F(a,a-b;a-b+1;-1)+\frac{\Gamma(a-b)\Gamma(b+1)}{\Gamma(a)}\labell{iden}\eeqa
and the on-shell condition $4s+t+u=0$, one can write the final
result as Veneziano-type amplitude with the interplay between the
$t$
and $u$ channels, \beqa A^{PR_2}&=&2(A_1+A_2)\nonumber\\
&=&\frac{i\kappa\,
T'_p}{4}\left(a_3B(-1-t/2,1-u/2)\right.\nonumber\\
&&\qquad\qquad\qquad \left.+ a_1B(-t/2,-u/2)+
a_2B(1-t/2,-u/2)\frac{}{}\right)\labell{finalamp}\eeqa

To show that the amplitude satisfies the Ward identities, we
compare $A^{PR_2}$ with  $A^{D_2}$. Both amplitudes can be
rewritten as \beqa A^{D_2}&=&\frac{i\kappa\,
T_p}{2}\left(us\Tr(\pol_1\inn\pol_2^T)-2sa_1-\frac{t}{2}a_2\right)
\frac{\Gamma(-t/2)\Gamma(-2s)}{\Gamma(1-t/2-2s)}\nonumber\\
A^{PR_2}&=&\frac{i\kappa\,T'_p}{4}
\left(us\Tr(\pol_1\inn\pol_2^T)+2sa_1-\frac{t}{2}a_2\right)
\frac{\Gamma(-t/2)\Gamma(-u/2)}{\Gamma(1-t/2-u/2)}\labell{compare}\eeqa
Using the fact that $a_1^{PR_2}=-a_1^{D_2}$ and
$a_2^{PR_2}=a_2^{D_2}$, one realizes that the kinematic factors
are exactly the same. On the other hand, it has been  shown in
\cite{MG} that kinematic factor of disk amplitude satisfies the
Ward identities. Hence the projective plane amplitude satisfies
the ward identities too. Moreover, if one uses the graviton vertex
operators in any other picture, the final result would  be the
same as above. We will use the above amplitude in section 4 to
study the low energy effective action of O-planes. We note here
that the amplitude $A^{PR_2}$ is invariant under replacements
$(p_1,\pol_1,p_2,\pol_2)\rightarrow (p_1,\pol_1,p_2\inn
D,D\inn\pol_2^T\inn D)$, $(p_1,\pol_1,p_2,\pol_2)\rightarrow
(p_1\inn D,D\inn\pol_1^T\inn D,p_2,\pol_2)$, and
$(p_1,\pol_1,p_2,\pol_2)\rightarrow (p_1\inn D,D\inn\pol_1^T\inn
D,p_2\inn D, D\inn\pol_2^T\inn D)$.

\section{R-R boson amplitude} The R-R vertex operator in Type II superstring theory  in the
absence of O-plane is \cite{polch}\beq
V^{D_2}(\pol,p,z,\bz)=(P_-\,\slf_{i(n)})^{AB}\,\norm{V_{-1/2\,A}(p,z)}
\ \norm{\tV_{-1/2\,B}(p,\bz)}\ \ . \labell{vertspin} \eeq The
holomorphic components above are given by \beq
V_{-1/2\,A}(p,z)=e^{-\phi(z)/2}\,S_A(z)\,e^{ip\cdot X(z)}
\labell{vertspinor} \eeq and the antiholomorphic components have
the same form, but with the left-moving fields replaced by their
right-moving counterparts. We refer readers to \cite{MG} for our
conventions on $P_-$ and  $\slf_{i(n)}$.
Using the same discussion as for NS-NS vertex operator
\reef{ver2}, in the presence of O-plane, the R-R vertex operator
is given by \beqa
V^{PR_2}(\pol,p,z,\bz)&=&\frac{1}{2}(P_-\,\slf_{i(n)})^{AB}\,\left(\norm{V_{-1/2\,A}(p,z)}
\ \norm{\tV_{-1/2\,B}(p,\bz)}\right.
\labell{vertspin}\\
&&\left.+\left(\norm{V_{-1/2\,A}(p,z)}\norm{\tV_{-1/2\,B}(p,\bz)}\right)
_{\psi(z)\leftrightarrow D\cdot\tpsi(\bz),\,X(z)\leftrightarrow
D\cdot\tX(\bz),\,\phi(z)\leftrightarrow
\tphi(\bz)}\right)\nonumber \eeqa Now under replacement
$\psi(z)\leftrightarrow D\cdot\tpsi(\bz)$, one finds the following
replacement for the spin field \cite{polch1}: \beqa
\tS_A(\bz)\leftrightarrow (M)_A{}^B S_B(z)\labell{M}\eeqa The R-R
vertex operator \reef{vertspin} becomes\beqa
V^{PR_2}(\pol,p,z,\bz)&=&\frac{1}{2}(P_-\,\slf_{i(n)})^{AB}\,\left(\norm{e^{-\phi(z)/2}\,S_A(z)\,e^{ip\cdot
X(z)}} \ \norm{e^{-\tphi(\bz)/2}\,\tS_B(\bz)\,e^{ip\cdot
\tX(\bz)}}\right.
\labell{vertspin2}\\
&&\left.+\norm{e^{-\tphi(\bz)/2}\,(M^{-1})_A{}^D\tS_D(\bz)\,e^{ip\cdot
D\cdot \tX(\bz)}} \
\norm{e^{-\phi(z)/2}\,(M)_B{}^CS_C(z)\,e^{ip\cdot D\cdot
X(z)}}\right)\nonumber \eeqa Now one can work with the above
vertex operator and the propagators \reef{xcor} to calculate any
scattering amplitude. Alternatively, one may use the replacement
\cite{MG} \beqa \tX(\bz)\rightarrow D\inn
X(\bz)&,&\tpsi(\bz)\rightarrow
D\inn\psi(\bz)\,\,\,,\,\,\,\tphi(\bz)\rightarrow
\phi(\bz)\labell{repl}\eeqa to simplify  the world sheet
propagators. The matrix $D^{\mu\nu}$ in \reef{xcor} is  replaced
with $\eta^{\mu\nu}$, \ie \beqa \langle
X^{\mu}(z)\,X^{\nu}(\bw)\rangle&=&-\eta^{\mu \nu}\,\log(1+nz\bw)
\labell{xcor1}\\
\langle\psi^{\mu}(z)\,\psi^{\nu}(\bw)\rangle&=&-\frac{\sqrt{n}\eta^{\mu\nu}}{1+nz\bw}\,;\,\,\,\,\,\,
\langle\psi^{\mu}(\bw)\,\psi^{\nu}(z)\rangle\,=\,\frac{\sqrt{n}\eta^{\mu\nu}}{1+nz\bw}
\nonumber\nonumber\\
\langle\phi(z)\,\phi(\bw)\rangle&=&-\log(1+nz\bw)\ \ . \nonumber
\eeqa In the R-R vertex operator, one should first note that the
replacement \reef{repl} makes the following replacement for the
spin field \cite{MG}: \beq \tS_A(\bz)\rightarrow M_A{}^B\,
S_B(\bz) \labell{replaced} \eeq where matrix $M$ is the same
matrix that appears in \reef{M}. Hence, the R-R vertex operator
\reef{vertspin2} becomes\footnote{The replacement \reef{repl}
makes the NS-NS vertex operator \reef{ver2} to the form \beqa
V^{PR_2}(\pol,p,z,\bz)&\!\!\!=\!\!\!& \frac{1}{2}\left(\pol\inn
D\right)_{\mu\nu}\left(\norm{
V^{\mu}_{\alpha}(p,z)}\norm{V^{\nu}_{\beta}(p\inn D,\bz)}
+\norm{V^{\mu}_{\alpha}(p,\bz)}\norm{V^{\nu}_{\beta}(p\inn
D,z)}\right)\eeqa The first term above is the vertex operator in
the presence of D-brane. In general, the vertex operator in the
presence of O-plane can be written as\beqa
V^{PR_2}(\pol,p,z,\bz)&=&\frac{1}{2}\left(V^{D_2}(\pol,p,z,\bz)+V^{D_2}(\pol,p,\bz,z)\right)\eeqa}
 \beqa
V^{PR_2}(\pol,p,z,\bz)&=&\frac{1}{2}(P_-\,\slf_{i(n)}M)^{AB}\,\left(\norm{e^{-\phi(z)/2}\,S_A(z)\,e^{ip\cdot
X(z)}} \ \norm{e^{-\phi(\bz)/2}\,S_B(\bz)\,e^{ip\cdot D\cdot
X(\bz)}}\right.
\nonumber\\
&&\left.+\norm{e^{-\phi(\bz)/2}\,S_A(\bz)\,e^{ip\cdot X(\bz)}} \
\norm{e^{-\phi(z)/2}\,S_B(z)\,e^{ip\cdot D\cdot
X(z)}}\right)\labell{vertspin1} \eeqa For the special case that
$D^{\mu\nu}=\eta^{\mu\nu}$, one should recover the R-R vertex
operator of Type I theory which results from unorienting Type IIB
theory. In this case, the matrix $M=1$. Interchanging the
holomorphic part with the antiholomorphic part in the second line
above, one finds an extra minus sign which results from fermionic
nature of R vertex operator. Now using the relation
$(\slf_{(n)})^{AB}=-(-1)^{n(n+1)/2}(\slf_{(n)})^{BA}$ (see
Appendix of \cite{MG}), one finds zero result for $n=1,5$, and
non-zero result  for $n=3$ which is the only R-R vertex operator
surviving under unorienting Type IIB.

In the presence of D-brane and absence of O-plane, the term in the
first line of \reef{vertspin1} represents the R-R vertex operator.

The amplitude describing two R-R states scattering from
O-plane/D-brane can  be written as
\[
A\simeq \int d^2\!z_1\, d^2\!z_2\ \langle\,
V_1(p_1,\pol_1,z_1,\bz_1) \ V_2(p_2,\pol_2,z_2,\bz_2)\,\rangle
\]
where the R-R vertex operator appear in \reef{vertspin1}. The
amplitude can be separated as \beqa
A^{PR_2}&=&A_1+A_2+A_3+A_4\nonumber\\
A^{D_2}&=&A_1\nonumber\eeqa where \beqa
A_1&\simeq&\frac{1}{4}(P_-\,\slf_{1(n)}M)^{AB}(P_-\,\slf_{2(m)}M)^{CD}\int
d^2\!z_1\,d^2\!z_2\nonumber\\
&&\langle\norm{V_{-1/2\,A}(p_1,z_1)} \ \norm{V_{-1/2\,B}(p_1\inn
D,\bz_1)}\norm{V_{-1/2\,C}(p_2,z_2)} \ \norm{V_{-1/2\,D}(p_2\inn
D,\bz_2)}\rangle\nonumber\\
A_2&\simeq&\frac{1}{4}(P_-\,\slf_{1(n)}M)^{AB}(P_-\,\slf_{2(m)}M)^{CD}\int
d^2\!z_1\,d^2\!z_2\nonumber\\
&&\langle\norm{V_{-1/2\,A}(p_1,z_1)} \ \norm{V_{-1/2\,B}(p_1\inn
D,\bz_1)}\norm{V_{-1/2\,C}(p_2,\bz_2)} \ \norm{V_{-1/2\,D}(p_2\inn
D,z_2)}\rangle\nonumber\\
A_3&\simeq&\frac{1}{4}(P_-\,\slf_{1(n)}M)^{AB}(P_-\,\slf_{2(m)}M)^{CD}\int
d^2\!z_1\,d^2\!z_2\nonumber\\
&&\langle\norm{V_{-1/2\,A}(p_1,\bz_1)} \ \norm{V_{-1/2\,B}(p_1\inn
D,z_1)}\norm{V_{-1/2\,C}(p_2,z_2)} \ \norm{V_{-1/2\,D}(p_2\inn
D,\bz_2)}\rangle\nonumber\\
A_4&\simeq&\frac{1}{4}(P_-\,\slf_{1(n)}M)^{AB}(P_-\,\slf_{2(m)}M)^{CD}\int
d^2\!z_1\,d^2\!z_2\nonumber\\
&&\langle\norm{V_{-1/2\,A}(p_1,\bz_1)} \ \norm{V_{-1/2\,B}(p_1\inn
D,z_1)}\norm{V_{-1/2\,C}(p_2,\bz_2)} \ \norm{V_{-1/2\,D}(p_2\inn
D,z_2)}\rangle\nonumber \eeqa The correlation function   for
superghost in $A_1$ is \beqa \langle\norm{e^{-\phi(z_1)/2}}
\norm{e^{-\phi(\bz_1)/2}}\norm{e^{-\phi(z_2)/2}}
\norm{e^{-\phi(\bz_2)/2}}\rangle&\!\!\!\!\!=\!\!\!\!\!&
\left((1+nz_1\bz_1)|z_1-z_2|^2|1+nz_1\bz_2|^2(1+nz_2\bz_2)\right)^{-1/4}\nonumber\eeqa
Now to find the correlation function between spin fields, we use
the following result \cite{fms}: \beqa \langle
:S_A(z_1)::S_B(z_2)::S_C(z_3)::S_D(z_4):\rangle&=&\frac{z_{14}z_{23}(\gamma^{\mu})_{AB}(\gamma_{\mu})_{CD}
-z_{12}z_{34}(\gamma^{\mu})_{AD}(\gamma_{\mu})_{BC}}{2(z_{12}z_{13}z_{14}z_{23}z_{24}z_{34})^{3/4}}\nonumber\eeqa
where $z_{ij}=z_i-z_j$. In the above relation the propagator of
world-sheet fermion is
$\langle\psi^{\mu}(z_i)\psi^{\nu}(z_j)\rangle=-\eta^{\mu\nu}/(z_i-z_j)$.
In our case, the propagator is the same when both fields  are
holomorphic or antiholomorphic. However, when one field is
holomorphic and the other is antiholomorphic, the propagator is
given by \reef{xcor1}. Then, for the correlation function of spin
fields in $A_1$, one expects the following result:\beqa
-n\frac{|1+nz_1\bz_2|^2(\gamma^{\mu})_{AB}(\gamma_{\mu})_{CD}+(1+nz_1\bz_1)(1+nz_2\bz_2)
(\gamma^{\mu})_{AD}(\gamma_{\mu})_{BC}}{
2\left(-(1+nz_1\bz_1)|z_1-z_2|^2|1+nz_1\bz_2|^2(1+nz_2\bz_2)\right)^{3/4}}\eeqa
With this correlation function, one can easily verifies, after
performing the $X$-correlators, that the integrand in $A_1$ is
invariant under CKG of projective plane/disk. Fixing this symmetry
as before, one finds \beqa
A^{D_2}&\simeq&(P_-\,\slf_{1(n)}M)^{AB}(P_-\,\slf_{2(m)}M)^{CD}\left(\frac{1}{2}
(\gamma^{\mu})_{AB}(\gamma_{\mu})_{CD}B(-t/2,-2s)\right.\nonumber\\
&&\left.\qquad\qquad\qquad+\frac{1}{2}
(\gamma^{\mu})_{AD}(\gamma_{\mu})_{BC}B(-t/2,1-2s)\right)\nonumber\\
A^{PR_2}_1&\simeq&-\frac{1}{4}(P_-\,\slf_{1(n)}M)^{AB}(P_-\,\slf_{2(m)}M)^{CD}\left(\frac{1}{2}
(\gamma^{\mu})_{AB}(\gamma_{\mu})_{CD}\frac{F(1+2s,-t/2;1-t/2;-1)}{-t/2}\right.\nonumber\\
&&\left.\qquad\qquad\qquad+\frac{1}{2}
(\gamma^{\mu})_{AD}(\gamma_{\mu})_{BC}\frac{F(2s,-t/2;1-t/2;-1)}{-t/2}\right)
\eeqa The disk amplitude is the one that has been found in
\cite{igorb,MG}. Both amplitude have the same massless pole,
however, the overall sign is different. This is consistent with
the observation  that  the massless closed string  couplings to
D-brane and to O-plane have different sign.

Similar calculation for $A_2$, $A_3$, and $A_4$  gives the
following results:\beqa
A^{PR_2}_2&\!\!\!\!\simeq\!\!\!\!&-\frac{1}{4}(P_-\,\slf_{1(n)}M)^{AB}(P_-\,\slf_{2(m)}M)^{CD}\left(\frac{1}{2}
(\gamma^{\mu})_{AB}(\gamma_{\mu})_{CD}\frac{F(1+2s,1-u/2;2-u/2;-1)}{1-u/2}\right.\nonumber\\
&&\left.\qquad\qquad\qquad+\frac{1}{2}
(\gamma^{\mu})_{AD}(\gamma_{\mu})_{BC}\frac{F(2s,-u/2;1-u/2;-1)}{-u/2}\right)\nonumber\\
A^{PR_2}_3&=&A^{PR_2}_2\nonumber\\
A^{PR_2}_4&=&A^{PR_2}_1
 \eeqa  Now summing all contributions and using the identity \reef{iden}, one finds
 \beqa
 A^{PR_2}&\simeq&-\frac{1}{2}(P_-\,\slf_{1(n)}M)^{AB}(P_-\,\slf_{2(m)}M)^{CD}\left(\frac{1}{2}
(\gamma^{\mu})_{AB}(\gamma_{\mu})_{CD}B(-t/2,1-u/2)\right.\nonumber\\
&&\left.\qquad\qquad\qquad+\frac{1}{2}
(\gamma^{\mu})_{AD}(\gamma_{\mu})_{BC}B(-t/2,-u/2)\right)\eeqa In
this case, like \reef{compare}, the $PR_2$ and $D_2$  amplitudes
have the same kinematic factor, \ie \beqa A^{D_2}&\simeq
&K(1,2)\frac{\G(-t/2)\G(-2s)}{\G(1-t/2-2s)}\nonumber\\
A^{PR_2}&\simeq
&\frac{1}{2}K(1,2)\frac{\G(-t/2)\G(-u/2)}{\G(1-t/2-u/2)}\labell{aa2}
\eeqa where the kinematic factor is \beqa
K(1,2)&=&\frac{u}{4}\Tr(P_-\,\slf_{1(n)}M\g^{\mu})\Tr(P_-\,\slf_{2(m)}M\g_{\mu})
+s\Tr(P_-\,\slf_{1(n)}M\g^{\mu}\slf_{2(m)}M\g_{\mu})\eeqa To write
the kinematic factor in terms of only momenta and polarization
tensors, one must perform the traces over gamma matrices which for
general $n$, $m$, $p$ is rather complicated--see\cite{igorb}.


\section{Low energy effective action} Now using the following
$\alpha'$ expansion for gamma function:\beqa
\frac{\G(-\alpha'a)\G(-\alpha'b)}{\G(1-\alpha'a-\alpha'b)}&=&
\frac{1}{\alpha'^2ab}-\frac{\pi^2}{6}+O(\alpha')\eeqa one finds
the low energy expansion for the scattering amplitudes
\reef{final}. In the case of NS-NS amplitude, one finds \beqa
A_{D_2}&=&\frac{i\kappa
T_p}{2}\left(-\Tr(\pol_1\inn\pol_2^T)-\frac{1}{t}\left(4s\Tr(\pol_1\inn\pol_2^T)+
2a_1\right)-\frac{1}{s}\left(\frac{a_2}{2}\right)\right.\nonumber\\
&&\left.+\frac{\pi^2}{12}\left(2s(t+4s)\Tr(\pol_1\inn\pol_2^T)
+4sa_1+ta_2\right)+\cdots\right)\labell{expand}\\
A_{PR_2}&=&\frac{i\kappa
T'_p}{4}\left(\frac{1}{t}\left(4s\Tr(\pol_1\inn\pol_2^T)-
2a_1\right)-\frac{2}{u}\left(a_1+a_2\right)\right.\nonumber\\
&&\left.+\frac{\pi^2}{12}\left(2s(t+4s)\Tr(\pol_1\inn\pol_2^T)
-4sa_1+ta_2\right) +\cdots\right)\nonumber\eeqa For $D^2$ case, it
has been shown that the massless poles are reproduced by field
theory which includes supergravity action for the bulk and the DBI
action, \beqa S^{D_2}_0&=&-T_p\int d^{p+1}x\, e^{-\Phi}\sqrt{-\det
(g_{ab}+B_{ab})}\nonumber\eeqa
 for D-brane action \cite{MG}. In particular, the $t$-channel in field theory for graviton-graviton amplitude is
\beqa A_{t}^{gg}(p_1,\pol_1,p_2,\pol_2)&=&-\frac{i\k
}{t}T_p\Big[\frac{}{}2p_1\inn \pol_2\inn\pol_1\inn p_2+
2p_2\inn\pol_1\inn\pol_2\inn D \inn p_2
\nonumber\\
&&+2p_1\inn\pol_2\inn\pol_1\inn D \inn p_1+2p_1\inn\pol_2\inn
D\inn\pol_1\inn p_2-p_1\inn D\inn p_1 \Tr(\pol_1\inn\pol_2)
\nonumber\\
&&-\Tr(\pol_1\inn D)\,p_1\inn\pol_2\inn p_1-\Tr(\pol_2\inn
D)\,p_2\inn\pol_1\inn p_2\Big]
\nonumber\\
&&-i\k T_p\Big[\Tr(\pol_1\inn\pol_2)+\Tr(D\inn\pol_1\inn\pol_2)
\vphantom{\frac{}{}}\Big]\ \ \labell{hhampone} \eeqa which is the
same as the massless $t$-channel in $A^{D_2}$ up to some
$(\alpha')^0$ contact terms. Moreover, the $(\alpha')^0$ contact
terms are exactly the ones that are reproduced by the above DBI
action. This conforms that the constant $T_p$ in the D-brane
amplitude \reef{norm} is tension of D-brane. The $(\alpha')^2$
contact terms, on the other hand, are shown to be reproduced by
the following action \cite{CPB}: \beqa
S^{D_2}_2&\!\!\!=\!\!\!&\frac{\pi^2\alpha'^2}{48}T_p\,
e^{-\Phi}\sqrt{-\det{g_{ab}}}
\left(R_{abcd}R^{abcd}-2R_{ab}R^{ab}-R_{ijab}R^{ijab}+2R_{ij}R^{ij}\right)\labell{rtwo}\eeqa

For $PR_2$ case, the $(\alpha')^0$ order of O-plane action should
be \cite{polch1}
 \beqa
S^{PR_2}_0&=&T'_p\int d^{p+1}x e^{-\Phi}\sqrt{-\det
g_{ab}}\labell{pr20}\eeqa where we take tension of O-plane to be
negative, \ie  $T'_p$ is absolute value of O-plane tension. To
check that the massless poles are reproduced by the effective
actions, we recall that the wave function for gravitons are given
by \reef{polg}. So the $t$-channel of scattering of two gravitons
is \beqa
A_t^{PR_2}(g,g)&=&\frac{T'_p}{-T_p}\left(A_t^{gg}(p_1,\pol_1/2,p_2,\pol_2/2)+A_t^{gg}(p_1\inn
D,D\inn\pol_1\inn D/2,p_2\inn D,D\inn\pol_2\inn
D/2)\right)\nonumber\\
&=&\left(\frac{1}{2}\right)\left(\frac{T'_p}{-T_p}\right)A_t^{gg}(p_1,\pol_1,p_2,\pol_2)\nonumber\eeqa
One can check that the above massless pole  and the massless
$t$-channel in \reef{expand} are the same up to some $(\alpha')^0$
contact terms. The $u$-channel can be read from \reef{hhampone} by
the replacement $(p_1,\pol_1,p_2,\pol_2)\rightarrow
(p_1,\pol_1/2,p_2\inn D, D\inn \pol_2\inn D/2)$ and
$(p_1,\pol_1,p_2,\pol_2)\rightarrow (p_1\inn D ,D\inn\pol_1\inn
D/2,p_2,  \pol_2/2)$, \ie \beqa A_u^{PR_2}(g,g)&=&
\frac{T'_p}{-T_p}\left(A_t^{gg}(p_1,\pol_1/2,p_2\inn
D,D\inn\pol_2\inn D /2)+A_t^{gg}(p_1\inn D,D\inn\pol_2\inn
D/2,p_2,\pol_2/2)\right)\nonumber\\
&=&\left(\frac{1}{2}\right)\left(\frac{T'_p}{-T_p}\right)A_t^{gg}(p_1,\pol_1,p_2\inn
D,D\inn\pol_2\inn D )
\eeqa One can again check that the above massless pole and the
massless $u$-channel pole in \reef{expand} are exactly the same up
to some $(\alpha')^0$ contact terms. To find these contact terms,
one should subtract the field theory amplitudes
$A_t^{PR_2}+A_u^{PR_2}$ from string theory amplitude
\reef{expand}, \ie \beqa
A_{PR_2}(g,g)-A_t^{PR_2}(g,g)-A_u^{PR_2}(g,g)&=&\frac{i\k
T'_p}{2}\left(\frac{1}{2}\Tr(\pol_1\inn D)\Tr(\pol_2\inn
D)-\Tr(\pol_1\inn\pol_2)\right.\nonumber\\
&&-\Tr(\pol_1\inn D\inn\pol_2\inn D)-2\Tr(\pol_1\inn
D\inn\pol_2)\labell{pr22}\\
&&\left.+\frac{\pi^2}{24}\left(2s(t+4s)\Tr(\pol_1\inn\pol_2)
-4sa_1+ta_2\right) +\cdots\right)\nonumber\eeqa The above
amplitude is symmetric under replacements
$(p_1,\pol_1,p_2,\pol_2)\rightarrow (p_1,\pol_1,p_2\inn
D,D\inn\pol_2\inn D)$, $(p_1,\pol_1,p_2,\pol_2)\rightarrow
(p_1\inn D,D\inn\pol_1\inn D,p_2,\pol_2)$, and
$(p_1,\pol_1,p_2,\pol_2)\rightarrow (p_1\inn D,D\inn\pol_1\inn
D,p_2\inn D, D\inn\pol_2\inn D)$.
 So instead of considering wave function \reef{polg}, one can
consider \reef{waved}  to find the effective action consistent
with the above couplings.  The two graviton couplings from
\reef{pr20} is exactly the $(\alpha')^0$ contact terms in the two
lines above. This confirms  that the tension of O-plane is
negative, \ie $-T'_p$.

It was shown in \cite{CPB} that the two gravitons couplings from
action \reef{rtwo} with the wave function \reef{waved} are exactly
equal to the $(\alpha')^2$ contact terms of  $A^{D_2}$, \eg the
terms in the second line of \reef{expand}.  On the other hand the
$(\alpha')^2$ contact terms of $A^{PR_2}$ are the same as the
$(\alpha')^2$ contact terms of $A^{D_2}$ up to overall factor.
Therefore, the action which is consistent with $(\alpha')^2$
contact terms of $A^{PR_2}$, \eg the terms in the third line of
\reef{pr22} is \beqa S^{PR_2}_2({\rm
gravity})&=&\left(\frac{1}{2}\right)\left(\frac{T'_p}{T_p}\right)S^{D_2}_2({\rm
gravity})\labell{rtwo2}\eeqa The same factor of
$T'_p/2T_p=2^{p-6}$, where we have used the relation between the
O-plane and D-brane  in Type II superstring theory  \cite{polch},
appears in the $R^2$ couplings in Wess-Zumino part of O-plane
effective action \cite{KDD}. The above action has been also found
in \cite{HJS}.

Now we would like to show that the same factor appears in the
Kalb-Ramond couplings at $(\alpha')^2$ order. To see this,
consider the scattering of two antisymmetric Kalb-Ramond tensors
from O-plane. In field theory, the scattering amplitude with wave
function \reef{waved} is given by
 \beqa
A_{t}^{BB}(p_1,\pol_1,p_2,\pol_2)&=&\frac{i\k}{t}T'_p\Big[2p_1\inn
\pol_2\inn\pol_1\inn p_2+p_1\inn D\inn p_1\Tr(\pol_1\inn\pol_2)
\nonumber\\
&&+2\left(\frac{}{}p_1\inn D\inn\pol_2\inn\pol_1\inn
p_2+p_1\inn\pol_2\inn D\inn\pol_1\inn p_2+ p_2\inn
D\inn\pol_1\inn\pol_2\inn p_1\right)
 \Big]
\nonumber\\
&&i\k T'_p\Tr(D\inn\pol_1\inn\pol_2) \,\, \labell{ampBB} \eeqa The
$t$-channel amplitude in field theory is\beqa
A_t^{PR_2}(B,B)&=&A_t^{BB}(p_1,\pol_1/2,p_2,\pol_2/2)+A_t^{BB}(p_1\inn
D,-D\inn\pol_1\inn D/2,p_2\inn D,-D\inn\pol_2\inn
D/2)\nonumber\\
&=&\frac{1}{2}A_{t}^{BB}(p_1,\pol_1,p_2,\pol_2)\eeqa The
$u$-channel amplitude in field theory is \beqa
A_u^{PR_2}(B,B)&=&A_t^{BB}(p_1,\pol_1/2,p_2\inn D,-D\inn\pol_2\inn
D/2)+A_t^{BB}(p_1\inn D,-D\inn\pol_1\inn D/2,p_2,\pol_2/2)\nonumber\\
&=&\frac{1}{2}A_t^{BB}(p_1,\pol_1,p_2\inn D,-D\inn\pol_2\inn
D)\labell{ampBBone}
\eeqa These poles are exactly the ones appear in $A^{PR_2}$.
Subtracting them, one finds
 \beqa
A_{PR_2}(B,B)-A_t^{PR_2}(B,B)-A_u^{PR_2}(B,B)&\!\!\!=\!\!\!&\frac{i\k
T'_p}{2}\left(\frac{\pi^2}{24}\left(2su\Tr(\pol_1\inn\pol_2)
-4sa_1+ta_2\right)\right.\nonumber\\
&&\left.\qquad\qquad\qquad\qquad
+\cdots\frac{}{}\right)\labell{pr222}\eeqa The above amplitude is
invariant under replacement  $(p_1,\pol_1,p_2,\pol_2)\rightarrow
(p_1,\pol_1,p_2\inn D,-D\inn\pol_2\inn D)$,
$(p_1,\pol_1,p_2,\pol_2)\rightarrow (p_1\inn D,-D\inn\pol_1\inn
D,p_2,\pol_2)$, and $(p_1,\pol_1,p_2,\pol_2)\rightarrow (p_1\inn
D,-D\inn\pol_1\inn D,p_2\inn D,-D\inn\pol_2\inn D)$. Note that
there is no $(\alpha)^0$ contact terms in the above amplitude
which is consistent with the $(\alpha')^0$ order action of O-plane
\reef{pr20} which has no $B$-field.  On other hand, the
$(\alpha')^2$ contact terms  are the same as the $(\alpha')^2$
contact terms of $A_{D_2}(B,B)$ in \reef{expand} up to the overall
factor. Hence, the $(\alpha')^2$ action of O-plane is proportional
to the action of D-brane, \ie \beqa S^{PR_2}_2({
B,B})&=&\left(\frac{1}{2}\right)\left(\frac{T'_p}{T_p}\right)S^{D_2}_2({
B,B})\eeqa An $(\alpha')^2$ correction to the  D-brane action,
$S^{D_2}_2({ B,B})$, has been proposed  in \cite{HJS}. One expects
 the above correspondence
between couplings to O-plane and to D-brane should be hold  for
other massless fields. However, at higher $\alpha'$ order, the
contact terms of $A^{PR_2}$ and $A^{D_2}$ are not proportional to
each other. So their effective actions are not proportional to
each other either.

\section{Discussion}

In this paper, we have presented detailed calculations of
two-point amplitudes describing massless NS-NS and RR closed
string  scattering from an Orientifold $p$-plane in ten
dimensions. Using the result for NS-NS amplitude,  we have derived
the effective action of O-plane up to order $(\alpha')^2$. While
the action at $(\alpha')^0$ order is consistent with the negative
cosmological constant term of O-plane, the couplings at
$(\alpha')^2$ are the same as the corresponding coupling in the
effective action of D-brane with relative factor
$1/2\left(T'_p/T_p\right)=2^{p-6}$.

We have seen that the R-R amplitude \reef{aa2} and NS-NS amplitude
\reef{compare} can be written as \beqa
A^{D_2}&=&\frac{i\k T_p}{2}K(1,2)\frac{\G(-t/2)\G(-2s)}{\G(1-t/2-2s)}\nonumber\\
A^{PR_2}&=&\frac{i\k
T'_p}{4}K(1,2)\frac{\G(-t/2)\G(-u/2)}{\G(1-t/2-u/2)}\labell{final}\eeqa
where the kinematic factor $K(1,2)$ is exactly the same in both
amplitudes. This form of amplitude in general is consistent with
the fact that the massless $t$-channel pole in  O-plane amplitude
is proportional to minus  massless $t$-channel pole in D-brane
amplitude. Hence,  we speculate  that the final result for any
other massless amplitude should also be written as in \reef{final}
with the appropriate kinematic factor that  has been  found in
\cite{MG} in the case of scattering from D-brane.

We have seen that  our calculation can not fix the the value of
constants $T'_p$ and $T_p$ in amplitude \reef{norm}. However, it
indicates that the coupling of massless closed string to O-plane
is given by $-T'_p$, and to D-brane by $T_p$. On the other hand,
as we have clarified  in the paragraph under \reef{norm} the two
constants $T_p'$ and $T_p$ have the same sign. Using the fact that
the coupling of graviton to an extended object is given by the
tension of the object, this indicates  that  the $PR_2$
calculations fix the tension of O-plane to be negative. This is
unlike the derivation of O-plane tension from interaction of
O-plane with D-brane which gives both positive and negative result
for the tension of O-plane, \ie $T'_p=\mp2^{p-5}T_p$, depending on
the Chan-Paton factor of the open string stretching between
D-brane and its image \cite{Dbreview}.

We have seen that the massless closed string vertex operators in
the presence of O-plane must be in the form of \reef{ver2} in
order to have gauge invariant scattering amplitude. This vertex
includes  the usual vertex operator of Type II theory and its
image. Using this vertex, one can calculate the coupling of this
state with a massless NS open string on the world volume of a
D-brane \cite{MRG}. The result is zero. Alternatively,  one can
calculate the scattering amplitude of NS-NS massless closed string
from a D-brane. It is given by \reef{finone} and another terms
which can be read from \reef{finone} by replacements
$p_2\rightarrow p_2\inn D$ and $\pol_2\rightarrow
D\inn\pol_2^T\inn D$. One can easily find the massless open string
pole is cancelled in the final amplitude. All these are consistent
with the fact that a D-brane at the fixed plane is unoriented.
However, in the scattering of external states from D-brane at disk
level, one can not fix the position of D-brane. Hence, in the
above amplitude the D-brane is not necessary to be   coincident
with O-plane. One should not conclude  that this  D-brane  is also
unoriented. If a D-plane is not coincident on O-plane then there
is  another D-brane which is image of the first one. In the
scattering of closed string from these branes at disk level, the
off-shell open strings see both D-branes to be coincident. Hence
the  open strings see only two unoriented D-branes at the position
of O-plane.

We have discussed the massless poles of  the string scattering
amplitudes in previous section. As well as these poles, the string
amplitudes contain an infinite number of massive poles as well.
 Each of the higher
 poles represents a massive closed string state coupling to the
O$p$-plane. From this point of view, the O$p$-plane provides a
$\delta$-function source for each of these massive states, just as
it does for the massless states. The same set of
$\delta$-functions in the transverse coordinates appears in the
construction of a boundary state description of a O-plane
\cite{CC,igorc}. This would then lend itself to an interpretation
of O-planes as objects of zero thickness. However, since O-plane
is not only a source of the massless fields but also massive
fields with $m^2=4n/\alpha'$, the conventional (low energy)
spacetime picture will breakdown at distances of the order of
$\sqrt{\alpha'}$. It is within this range that the full closed
string spectrum makes its presence felt. Hence from this
perspective, one would ascribe a thickness of the order of
$\sqrt{\alpha'}$ to O-planes.

An exception to the above form of the amplitudes is the special
case of $p=8$. In this case, the O$_8$-plane forms a domain wall
dividing the ten-dimensional spacetime into two halves. The
transverse space is only one-dimensional running along $x^9$, and
so it is only the ninth component of the momentum vectors which is
not conserved. Hence, \beqa s=-(p_1^9)^2=-(p_2^9)^2\eeqa Since
there is a single component of momentum appearing here, one has
either $p_1^9=-p_2^9$ or $p_1^9=p_2^9$. In the first case, one
finds $t=0$, and in the second case one finds $u=0$.  This result
allows one to rewrite the amplitude \reef{compare} in a
drastically simple form \beqa A_{PR_2}&=&\frac{i\kappa
T'_p}{4}\left(\frac{1}{t}\left(4s\Tr(\pol_1\inn\pol_2^T)-
2a_1\right)-\frac{2}{u}\left(a_1+a_2\right)\right)\eeqa As we have
seen in the previous section, this is exactly the amplitude that
one finds in the low energy limit. Hence, all massive poles
disappear in O$_8$-plane amplitudes. Similar disappearance  of
massive states happens in the $D_{-1}$-brane amplitude
\cite{igora,igorb,MG}.

While there are no massive poles in the above amplitude, one
should not conclude that no massive string states couple to
O$_8$-plane. The coupling of massive states to external massless
states are such that in the O$_8$-plane amplitude all massive
poles disappear. One expects then the massive poles make their
appearance in the scattering amplitudes involving massive external
strings. Similar discussion can be used for $D_{-1}$-brane case
\cite{MG}. To clarify this point, one may consider bosonic string
theory and work with massless and tachyon states. The scattering
amplitude of two tachyons with O$_p$-plane can be found to be\beqa
A^{PR_2}&\simeq&\frac{\G(-1-u/2)\G(-1-t/2)}{\G(-2-u/2-t/2)}\nonumber\eeqa
For $p=24$, this simplifies to \beqa
A^{PR_2}&\simeq&\frac{1}{-u/2}+\frac{1}{-t/2}\nonumber\eeqa where
only massless poles survives. Now consider the scattering
amplitude of one tachyon and one graviton with O$_{p}$-plane. The
result is  \beqa A^{PR_2}&\!\!\!\simeq\!\!\!& \Tr(\pol_2\inn
D)\frac{\G(-u/2)\G(-t/2)}{\G(-u/2-t/2)}+p_2\inn D\inn\pol_2\inn
D\inn p_2\frac{\G(1-u/2)\G(-1-t/2)}{\G(-u/2-t/2)}\labell{tg}\\
&&+2p_2\inn D\inn\pol_2\inn D\inn
p_1\frac{\G(-u/2)\G(-1-t/2)}{\G(-1-u/2-t/2)}+p_1\inn
D\inn\pol_2\inn D\inn
p_1\frac{\G(-1-u/2)\G(-1-t/2)}{\G(-2-u/2-t/2)}\nonumber\eeqa For
the special case of $p=24$, the amplitude for $u=0$ case
simplifies to \beqa A^{PR_2}&\simeq&\Tr(\pol_2\inn
D)\left(\frac{1}{-t/2}+\frac{1}{-u/2}\right)+\frac{p_1\inn\pol_2\inn
p_1}{-1-t/2}\nonumber\eeqa which has both massless and tachyon
poles. If one consider other external massive states, one would
find other massive poles.  Hence one concludes that all massive
states couple to the domain wall O-planes  like any other
O$_p$-plane. This should also be evident given the complex form of
the boundary states describing a O$_8$-plane, and  the complex
form of the potential energy between an O$_8$-plane and a
D$_8$-brane \cite{Dbreview}. Hence, domain wall O-planes also have
thickness of order $\sqrt{\alpha'}$.

\section*{Acknowledgments}
I gratefully acknowledge useful conversations with A. Sen, M.M.
Sheikh-Jabbari and A. Ghodsi. I would  also like to thank H.
Ghorbani for double checking the result in  equation 20.


\begin{thebibliography}{99}


\bibitem
{joe}{J.~Polchinski, ``Dirichlet-Branes and Ramond-Ramond
Charges'', arXiv: hep-th/9510017.}
\bibitem
{Dbreview}{J.~Polchinski, S.~Chaudhuri and C.V.~Johnson, ``Notes
on D-Branes,'' arXiv: hep-th/9602052.}
\bibitem
{KD}{ B. Craps and F. Roose, Phys. Lett. B {\bf 445}, 150 (1998)
[arXiv:
hep-th/9808074];\\
B. Stefanski, Jr., Nucl. Phys. B {\bf 548}, 275 (1999) [arXiv:
hep-th/9812088];\\
J. A. Scrucca, and M. Serone, Nucl. Phys. B {\bf 552}, 291 (1999)
[arXiv: hep-th/9812071].}
\bibitem
{HJS}{H. J. Schnitzer, and N. Wyllard, JHEP {\bf 0208}, 012 (2002)
[arXiv: hep-th/0206071]}
\bibitem
{Dbrane}{J.~Dai, R.G.~Leigh and J.~Polchinski, Mod. Phys. Lett.
{\bf A4} (1989) 2073; J.~Polchinski, Phys. Rev. {\bf D50}
(1994) 6041.}







\bibitem
{igora}{I.R.~Klebanov and L.~Thorlacius, ``The Size of p-Brane'',
arXiv: hep-th/9510200.}

\bibitem
{igorb}{S.S.~Gubser, A.~Hashimoto, I.R.~Klebanov and
J.M.~Maldacena, Nucl. Physics. B {\bf 472}, 231 (1996) arXiv:
hep-th/9601057.}

\bibitem
{form}{J.L.F.~Barbon, Phys. Lett. B {\bf 382}, 60 (1996) [arXiv:
hep-th/960198].}
\bibitem
{MG}{M. R. Garousi, and R. C. Myers, Nucl. Phys. B {\bf 475}, 193
(1996) [arXiv: hep-th/9603194].}
\bibitem
{CC}{C. Callan, C. Lovelace, C. Nappi and S. Yost, Nucl. Phys. B
{\bf 293}, 83 (1987).}


\bibitem
{polch}{J. Polchinski, ``TASI Lectures on D-branes'' arXiv:
hep-th/9611050.}
\bibitem
{danf}{D. Friedan, ``Notes on String Theory and Two Dimensional
Conformal Field Theory,'' in {\em Unified String Theories,} Eds.
M.B.~Green and D.J.~Gross (World Scientific Publishing, 1986).}
\bibitem
{polch1}{J. Polchinski, {\em String theory}, Vol I and II
(Cambridge Univ. Press, Cambridge, 1998).}
\bibitem
{MBG}{M. B. Green, and N. Seiberg, Nucl. Phys. B {\bf 299}, 559
(1988).}
\bibitem
{TH}{T. Huber and D. Maitre, Comput. Phys. Commun. {\bf 175}, 122
(2006) [arXiv: hep-ph/0507094].}
\bibitem
{gsw}{M.B.~Green, J.H.~Schwarz and E.~Witten, {\em Superstring
Theory} (Cambridge University Press, 1987).}

\bibitem
{wolf}{The wolfram function site, http://functions.wolfram.com/}
\bibitem
{fms}{D.~Friedan, S.~Shenker and E.~Martinec, Phys. Lett. {\bf
160B} (1985) 55; Nucl. Phys. {\bf B271} (1986) 93.}
\bibitem
{CPB}{C. P. Pachas, P. Bain, and M. B. Green, JHEP {\bf 05}, 011
(1999) [arXiv: hep-th/9903210].}
\bibitem
{KDD}{K. Dasgupta, D. P. Jatkar, and S. Mukhi, Nucl. Phys. B {\bf
523}, 465 (1998)  [arXiv: hep-th/9707224].}

\bibitem
{MRG} {M. R. Garousi and R. C. Myers, Nucl. Phys. B {\bf 542}, 73
(1999) [arXiv: hep-th/9809100].}

\bibitem
{igorc}{C.G.~Callan and I.R.~Klebanov, Nucl. Phys. B {\bf 465},
473 (1996) [arXiv: hep-th/9511173].}




\end{thebibliography}
\end{document}